\documentclass[numberedappendix]{emulateapj}

\usepackage{multirow}
\usepackage{graphicx}
\usepackage{natbib}
\usepackage{amsmath}

\newcommand{\vecv}{\mathbf{v}}

\newcommand{\vecT}{\mathbf{T}}
\newcommand{\calT}{\mathcal{T}}
\newcommand{\calG}{\mathcal{G}}
\newcommand{\calL}{\mathcal{L}}
\newcommand{\msun}{M_{\odot}}

\begin{document}

\title{On the Dynamics and Evolution of Gravitational Instability-Dominated Disks}

\author{Mark Krumholz}
\affil{Department of Astronomy \& Astrophysics, University of California, Santa 
Cruz, CA 95064 USA}
\email{krumholz@ucolick.org}

\author{Andreas Burkert}
\affil{University Observatory Munich (USM), Scheinerstrasse 1, 81679 Munich, Germany and Max-Planck-Institut fuer extraterrestrische Physik, Giessenbachstrasse 1, 85758, Germany, Max-Planck-Fellow}
\email{burkert@usm.uni-muenchen.de}

\begin{abstract}
We derive the evolution equations describing a thin axisymmetric disk of gas and stars with an arbitrary rotation curve that is kept in a state of marginal gravitational instability and energy equilibrium due to the balance between energy released by accretion and energy lost due to decay of turbulence. Rather than adopt a parameterized $\alpha$ prescription, we instead use the condition of marginal gravitational instability to self-consistently determine the position- and time-dependent transport rates. We show that there is a steady-state configuration for disks dominated by gravitational instability, and that this steady state persists even when star formation is taken into account if the accretion rate is sufficiently large. For disks in this state we analytically determine the velocity dispersion, surface density, and rates of mass and angular momentum transport as a function of the gas mass fraction, the rotation curve, and the rate of external accretion onto the disk edge. We show that disks that are initially out of steady state will evolve into it on the viscous timescale of the disk, which is comparable to the orbital period if the accretion rate is high. Finally, we discuss the implications of these results for the structure of disks in a broad range of environments, including high redshift galaxies, the outer gaseous disks of local galaxies, and accretion disks around protostars.
\end{abstract}


\section{Introduction}

In the past few years, observational and theoretical advances in many areas have led to intense study of the role of accretion and gravitational instability in determining the structure and rates of transport through disks. In the high redshift universe, clumpy star-forming galaxies at redshifts $z\sim 2-3$ \citep[e.g.][]{elmegreen04b, elmegreen05a, genzel08a, forster-schreiber09a, cresci09a} appear to be undergoing rapid accretion, and also have velocity dispersions that are much larger than those present in local galaxies. Numerical simulations \citep[e.g.][]{dekel09b, ceverino09b, agertz09a, bournaud09a, mcnally09a} suggest that the large velocity dispersion and the massive clump morphology are both produced by a combination of gravitational instability and rapid external accretion. Around active galactic nuclei, radiative cooling pushes thin accretion disks into a state of gravitational instability \citep{shlosman90a, goodman03a}, and in this state their accretion rates and structures may be determined by gravitationally-driven turbulence \citep{gammie01a}. Closer to home, accretion disks around protostars of mass $\ga 1$ $\msun$ are expected to experience strong gravitational instability for a significant part of their lives \citep{kratter08a} due to a combination of rapid accretion and strong radiative cooling. Numerical simulations indicate that the non-circular motions produced by this instability provide the dominant mechanism for mass and angular momentum transport in the disk \citep[e.g.][]{krumholz07a, krumholz09c}.

Dozens of simulations of gravitational instability in disks have been published, both for disks undergoing external accretion \citep[e.g.]{vorobyov07a, vorobyov08a, vorobyov09a, kratter10a, machida10a} and for those in isolation (e.g.\ \citealt{lodato04a, lodato05a, kim07, cai08a, cossins09a, agertz09a}; see \citealt{durisen07a} for a review of earlier work). Based on these, several authors have presented one-dimensional time-dependent disk evolution models in which the effects of gravitational instability are approximated by an $\alpha$ prescription, with $\alpha$ obtained by fits to simulation results or by general energy arguments \citep{goodman03a, hueso05a, kratter08a, rice09a, rice10a}. Similar order-of-magnitude energy arguments have been extended to the case of galactic disks by \citet{dekel09a}, \citet{klessen09b}, and \citet{elmegreen10a}. In the realm of purely analytic work, \citet{bertin99a} present steady-state solutions for self-gravitating disks with decaying turbulence, while \citet{rafikov09a} and \citet{clarke09a} derived steady-state accretion rates for disks in balance between radiative cooling and accretion-driven heating in protostellar disks.

While the analytic and one-dimensional models have provided a good understanding of the basic mechanism of gravitationally-driven turbulence and transport in disks, they also suffer from significant weaknesses. No analytic models published to date consider the case of gravitational instability-dominated disks that are time-dependent rather than in steady state. Most previous work has been limited to a particular rotation curve (e.g.\ Keplerian or flat), to a disk of pure gas without stars, and to a disk that is vertically supported by thermal pressure rather than supersonic turbulence. As a result of these limitations it is not clear under what circumstances disks that are not in equilibrium can be expected to evolve into it, and it is not even clear what the equilibrium state is for a star-forming, supersonically turbulent disk such as a galactic disk.

Our goal is to improve this situation by developing a first-principles theory for the evolution of a thin, supersonically turbulent disk of star-forming gas in a state of marginal gravitational stability, driven by a specified rate of external accretion. The only assumptions we make are (1) that the disk maintains a state of marginal gravitational instability ($Q=1$) at all times, and (2) the rate of energy loss due to radiative cooling can be parameterized as a certain fraction of the energy per crossing time of a disk scale height. We do not assume that the disk is in steady state or that it is characterized by any particular rotation curve. From these assumptions, in Section \ref{sec:eqns} we derive equations describing the instantaneous, position-dependent rates of mass, energy, and angular momentum transport, and the time evolution of the disk surface density and velocity dispersion. In Section \ref{sec:steady} we show that these equations admit an exact steady-state solution, and we derive the steady-state profiles of surface density, velocity dispersion, and transport of mass, energy, and angular momentum. In Section \ref{sec:nonsteady} we show that disks that are not in the steady state will evolve toward it, and that for high accretion rates this evolution occurs on an orbital timescale. Finally, in Section \ref{sec:discussion} we discuss the implications of our findings, and we summarize in Section \ref{sec:conclusion}.

\section{Evolution Equations}
\label{sec:eqns}

\subsection{Mass, Angular Momentum, and Energy Transport}

We begin from the equations describing the evolution of a viscous fluid in a gravitational field that is in the process of turning its mass into collisionless stars. These are the equation of continuity, the Navier-Stokes equation, and the first law of thermodynamics:
\begin{eqnarray}
\label{continuity}
\frac{1}{\rho} \frac{D\rho}{Dt} & = & -\nabla\cdot\vecv -\frac{\dot{\rho}_*}{\rho} \\
\label{navierstokes}
\rho \frac{D\vecv}{Dt} & = & - \nabla p - \rho\nabla\psi + \nabla\cdot\vecT \\
\label{firstlaw}
\rho \frac{De}{Dt} & = & -p\nabla\cdot \vecv + \Phi + \Gamma-\Lambda.
\end{eqnarray}
Here $\rho$, $\vecv$, $e$, and $p$, are the gas density, velocity, specific internal energy, and pressure, $\dot{\rho}_*$ is the rate per unit volume at which gas mass turns into stellar mass, $\psi$ is the gravitational potential, $\vecT$ is the viscous stress tensor, $\Phi=T^{ij}(\partial v_i/\partial x_j)$ is the dissipation function, and $\Gamma$ and $\Lambda$ are the volumetric rates of energy gain and loss due to non-fluid (e.g.\ radiative or chemical) processes. Note that no terms associated with star formation appear in the first law of thermodynamics or the Navier-Stokes equations because star formation does not alter the bulk velocity or specific internal energy of the gas.

%We also point out that, in the form in which we have chosen to express the dissipation function $\Phi$, the derivative of $\vecv$ is a coordinate derivative, {\it not} a covariant derivative. The distinction is significant because we will be working in cylindrical coordinates, where the two are not identical.

%MRK condensed
We consider a thin, axisymmetric disk centered on the origin lying in the plane $z=0$. At every radius $r$ the disk is characterized by a surface density $\Sigma$ and a total thermal plus non-thermal velocity dispersion $\sigma$. The material orbits the origin with angular velocity $v_\phi$ and has a radial velocity $v_r \ll v_\phi$. In Appendix \ref{derivation} we show that for such a star-forming disk equations (\ref{continuity}), (\ref{navierstokes}), and (\ref{firstlaw}) imply
\begin{eqnarray}
\label{eq:mcons}
\frac{\partial}{\partial t} \Sigma + \frac{1}{r}\frac{\partial}{\partial r}(r\Sigma v_r) & = & - \dot{\Sigma}_* \\
\label{eq:jcons}
\Sigma \left(\frac{\partial j}{\partial t} + v_r\frac{\partial}{\partial r}j\right) & = & \frac{1}{2\pi r} \frac{\partial}{\partial r} \calT \\
\lefteqn{\frac{1}{2}\Sigma \left[\frac{\partial}{\partial t}\left(v_{\phi}^2+3\sigma^2\right) + v_r\frac{\partial}{\partial r}\left(v_{\phi}^2+3\sigma^2+2\psi\right)\right] }
\qquad\qquad\qquad\qquad\qquad\quad
 &&
\nonumber \\
{}
+ \frac{1}{r} \frac{\partial}{\partial r} \left(r \Sigma v_r \sigma^2\right) &= &
\frac{1}{2 \pi r}\frac{\partial}{\partial r}(\Omega \calT) + \calG - \calL,
\label{eq:econs}
\end{eqnarray}
where 
\begin{equation}
\calT = 2\pi \int  r^2 T_{r\phi} \, dz
\end{equation}
is the viscous torque, $j=rv_\phi$ is the specific angular momentum, $\Omega=v_\phi/r$ is the angular velocity, $\beta = \partial \ln v_\phi / \partial \ln r$, and $\calG = \int \Gamma \, dz$ and $\calL=\int \Lambda \, dz$ are the vertically-integrated rates of non-fluid energy gain and loss. Equations (\ref{eq:mcons}), (\ref{eq:jcons}), and (\ref{eq:econs}) are the standard equations of mass, angular momentum, and energy conservation for a thin disk \citep[e.g.][]{balbus98a} generalized to the case of a supersonically turbulent gas that is forming stars.

If we assume that the disk is always close to radial force balance and that the potential varies slowly in time then we have $\partial\psi/\partial r \approx v_\phi^2/r$ and $\partial j/\partial t \approx 0$. Using these conditions in equations (\ref{eq:mcons}), (\ref{eq:jcons}), and (\ref{eq:econs}), we arrive at the evolution equations for $\Sigma$ and $\sigma$:
\begin{eqnarray}
\frac{\partial \Sigma}{\partial t} & = & \frac{1}{2\pi(\beta+1) r v_\phi} \left[\frac{\beta(\beta+1) + r\frac{\partial \beta}{\partial r}}{(\beta+1) r} \left(\frac{\partial \calT}{\partial r}\right) - \frac{\partial^2 \calT}{\partial r^2}\right] 
\nonumber \\
& & {} - \dot{\Sigma}_*
\label{masscons}
\\
\frac{\partial \sigma}{\partial t} & = &
\frac{\calG - \calL}{3\sigma \Sigma} + \frac{1}{6\pi r \Sigma}
\left[
(\beta-1) \frac{v_\phi}{r^2 \sigma} \calT
\right.
\nonumber \\
& &
{}
+\frac{\beta^2\sigma+ 
\sigma \left(r \frac{d\beta}{dr}+\beta\right) - 5 (\beta+1) r \frac{\partial\sigma}{\partial r}
}{(\beta+1)^2 r v_\phi}
\left(\frac{\partial\calT}{\partial r}\right)
\nonumber \\
& &
\left.
-\frac{\sigma}{(\beta+1)v_\phi} \left(\frac{\partial^2\calT}{\partial r^2}\right)
\right]
.
\label{veldisp}
\end{eqnarray}

Finally, we note that the underlying assumption of our model is that we can represent the transport processes in a disk dominated by gravitational instability using a local viscosity. This assumption is only valid in certain circumstances, and this sets limits on the applicability of our model that we discuss in \S~\ref{locality}.
%MRK end new section 2.1

\subsection{Star Formation and Radiative Gain and Loss}

Equations (\ref{masscons}) and (\ref{veldisp}) fully specify the time evolution of the system. The only physical approximations we have made thus far are that the disk is thin and axisymmetric, and that turbulent eddies on size scales of the disk scale height provide an effective pressure proportional to the square of the turbulent velocity dispersion. However, we have not yet determined the functions describing the star formation rate $\dot{\Sigma}_*$, the rates of radiative gain and loss $\calG$ and $\calL$, and the torque $\calT$. Since the physics involved in these terms is complex, we proceed in a simple parameterized manner.

For the star formation rate, we note that both observation \citep[e.g.][]{krumholz07e, bigiel08a, evans09a} and theory \citep[e.g.][]{krumholz05c, krumholz09b, padoan09a, hennebelle09a} indicate that molecular gas forms stars at a rate of $\sim 1\%$ of its mass per free-fall time. We compute the free-fall time using the mid-plane density; since the scale height is $H = \sigma/\Omega$, this is $\rho = \Sigma \Omega/\sigma$. Thus we adopt a star formation rate
\begin{equation}
\label{sfr}
\dot{\Sigma}_* = \epsilon_{\rm ff}\Sigma\sqrt{G\rho} = \epsilon_{\rm ff} \left(\frac{G \Sigma^3 \Omega}{\sigma}\right)^{1/2},
\end{equation}
where $\epsilon_{\rm ff} \approx 0.01$. The true value of $\epsilon_{\rm ff}$ is likely to be slightly higher than this, because the clouds where stars form are generally denser than the mean midplane density in the disk. Even with this correction, however, we can be confident that $\epsilon_{\rm ff} \la 0.1$. 
Moreover, if a significant fraction of the ISM is atomic rather than molecular, the star formation rate is greatly reduced \citep[e.g.][]{wyder09a, blanc09a}, in which case $\epsilon_{\rm ff}$ can be much smaller. More complex and accurate star formation laws are possible \citep[e.g.][]{krumholz09b}, but we will see below that such increased accuracy is not necessary at this stage. However, we do note that for a gaseous disk with $Q=1$ (see Section \ref{stability}, equation \ref{q1}) and a flat rotation curve ($\beta=0$), equation (\ref{sfr}) gives a star formation rate $\dot{\Sigma}_* = 0.0067 \Sigma \Omega$; in comparison, \citet{kennicutt98a} reports an observed star formation rate $\dot{\Sigma}_* = 0.011 \Sigma \Omega$, identical to within the errors.\footnote{The coefficient reported in \citet{kennicutt98a} is 0.017 rather than 0.011. The reduction to 0.011 comes from replacing the \citet{salpeter55} IMF used in Kennicutt's work to a \citet{chabrier05a} IMF.}

For the energy loss rate, we note that numerous simulations of turbulence show that it decays via radiative shocks on roughly a crossing timescale \citep[e.g.][]{stone98, maclow98}. For the purpose of estimating the loss rate we take the characteristic length scale of the turbulence to be comparable to the gas scale height $H$, so the crossing time is $1/\Omega$. This is consistent with observations that show that turbulent power is dominated by the largest scales. We also limit our attention to galaxies where the total velocity dispersion is significantly in excess of the thermal value, since these are the only galaxies where one needs to explain the observed velocity dispersion by appealing to physics other than radiative balance. Thus we take the kinetic energy per unit area to be $(3/2)\Sigma \sigma^2$. The condition that the disk lose this amount of energy per crossing time of the disk scale height then reduces to
\begin{equation}
\label{decayrate}
\calL = \eta \Sigma\sigma^2\Omega,
\end{equation}
where $\eta$ is a dimensionless number of order unity. As a fiducial parameter we adopt $\eta = 3/2$, which corresponds to radiating away the full kinetic energy every scale height-crossing time. In disks where the velocity dispersion is primarily thermal rather than non-thermal, the loss rate will assume a different functional form \citep[e.g.][]{rafikov09a}, but the remainder of our analysis will be unchanged.

The gain rate is much more complex, since it involves turbulent motions generated by star formation. Since we are interested in systems where gravitationally-driven turbulence dominates, however, we make the extreme assumption that $\calG = 0$. We return to the question of the real value of $\calG$ in Section \ref{sec:feedback}.

\subsection{Gravitational Stability and the Torque Equation}
\label{stability}

We now turn to the central hypothesis of our model, which is that the a self-gravitating disk will adjust its torque, and therefore its radial mass flux, so as to remain in a state of marginal stability. This hypothesis has also been investigated in the models of \citet{burkert09a}. For a disk of gas plus stars, the parameter that describes its stability is \citep{rafikov01}
\begin{equation}
Q(q)^{-1} = 2Q_*^{-1} \frac{1}{q} \left[1 - e^{-q^2}I_0(q^2)\right] + 2Q_g^{-1} R \frac{q}{1 + q^2 R^2},
\end{equation}
where
\begin{equation}
\label{q1}
Q_* = \frac{\kappa \sigma_*}{\pi G \Sigma_*}  \qquad
Q_g = \frac{\kappa \sigma}{\pi G \Sigma} \qquad
R = \frac{\sigma}{\sigma_*},
\end{equation}
and $I_0$ is the Bessel function of order zero. Here $\Sigma_*$ and $\sigma_*$ are the surface density and velocity dispersion of stars, $\kappa = [2(\beta+1)]^{1/2}\Omega$ is the epicyclic frequency, and $q=k \sigma_*/\kappa$ is the dimensionless wavenumber of the mode in question. Modes for which $Q(q) < 1$ are unstable. Note that \citet{romeo10a} have proposed a generalization of this condition for the case of gas with a scale-dependent velocity dispersion, as expected for turbulence, but for simplicity we use the \citet{rafikov01} criterion instead.

It is not generally possible to find the minimum value of $Q(q)$ analytically. However, we can obtain a significant simplification if we focus on the most interesting cases for gravitationally-driven turbulence. In disks at high redshift there has not been time for the stars and gas to evolve so that their velocity dispersions are very different -- see \S~\ref{sec:highz} for a further discussion of this point. For these disks we therefore adopt $\sigma_* = \sigma$. In this case the expression
\begin{equation}
\label{qapprox}
Q = \min(Q(q)) \approx \left(\frac{1}{Q_g} + \frac{1}{Q_*}\right)^{-1} = \frac{\kappa \sigma}{\pi G (\Sigma+\Sigma_*)}
\end{equation}
is accurate to better than 7\%. With this approximation, the condition for a disk to remain marginally stable at $Q=1$ becomes
\begin{eqnarray}
0 & = &
\frac{\partial Q}{\partial \sigma} \frac{\partial \sigma}{\partial t}
+ \frac{\partial Q}{\partial \Sigma} \frac{\partial \Sigma}{\partial t}
+ \frac{\partial Q}{\partial \Sigma_*} \frac{\partial \Sigma_*}{\partial t}\\
& = &
\frac{1}{\sigma}\frac{\partial\sigma}{\partial t}-\frac{1}{\Sigma+\Sigma_*}\left(\frac{\partial\Sigma}{\partial t} + \frac{\partial\Sigma_*}{\partial t}\right).
\label{qcondition}
\end{eqnarray}
Plugging in our expressions for the temporal derivatives of $\sigma$, $\Sigma$, and $\Sigma_*$, we obtain an equation that describes the torque required to maintain $Q=1$:
\begin{equation}
\label{torquedim}
f_2 \frac{\partial^2 \calT}{\partial r^2} + f_1 \frac{\partial \calT}{\partial r} + f_0 \calT =  F
\end{equation}
with
\begin{eqnarray}
f_0 & = & \left[\frac{\beta-1}{6\sqrt{2(\beta+1)}}\right] \frac{G}{f_g r^2 \sigma^3}
\\
f_1 & = & 
-\frac{(3 f_g-1) r \sigma \frac{d\beta}{dr} - (\beta+1)\left[(3 f_g-1) \beta\sigma+5 r \frac{\partial\sigma}{\partial r}\right]}
{6\sqrt{2(\beta+1)^5}}
\nonumber
\\
& & {}\times \frac{G}{f_g r v_\phi^2 \sigma}
 \\
f_2 & = & \left[\frac{3f_g-1}{6\sqrt{2(\beta+1)^3}}\right] \frac{G}{f_g v_\phi^2\sigma} \\
F & = & \frac{\eta v_\phi}{3r},
\end{eqnarray}
where $f_g = \Sigma/(\Sigma+\Sigma_*)$ is the gas fraction in the disk, and we have used $Q=1$ to replace any dependence on $\Sigma$ with a dependence on $\sigma$ and $f_g$. We refer to this as the torque equation.
Note that in deriving equation (\ref{torquedim}) we have implicitly assumed that stars do not migrate radially from their formation locations. If we were to make the opposite assumption, that stars and gas move together, then the combined gas plus star disk would act essentially like a purely gaseous one, except that the stars would be dissipation free. The corresponding torque equation is simply equation (\ref{torquedim}) with $\eta$ replaced by $\eta f_g$, and $f_g \rightarrow 1$ in all other terms.

The other interesting location to consider for gravitationally-driven turbulence is in outer parts of present-day disks. In these regions star formation occurs at a negligible rate, and $\sigma_* \gg \sigma$. In this case $Q \approx Q_g$, and a little calculation shows that the torque equation is simply (\ref{torquedim}) with $f_g = 1$ everywhere. The stars simply become irrelevant for gravitational stability. Thus equation (\ref{torquedim}), with the appropriate choice of $f_g$ and $\eta$, is capable of representing both a present-day outer galaxy disk and a high redshift disk.

To finish specifying the torque, we must provide boundary conditions for equation (\ref{torquedim}). One boundary condition comes from fixing the external accretion rate onto the galaxy to a value $\dot{M}=\dot{M}_{\rm ext}$ (so $\partial \calT/\partial r = -v_\phi(\beta+1) \dot{M}_{\rm ext}$) at the disk outer edge, $r=R$. We imagine the inflow rate at this radius to be set by cosmological infall, which occurs at a rate unaffected by what happens in the disk. The other boundary condition depends on how we handle the inner boundary. If we imagine that our model applies all the way to $r=0$, we must find a solution that remains regular in the vicinity of the singular point there. We show below that there does exist a steady-state solution that is regular at $r=0$ and has the specified accretion rate  $\dot{M}=\dot{M}_{\rm ext}$ at $r=R$. Outside of that steady state it is not possible to simultaneously have regularity at $r=0$ and an externally-imposed accretion rate at $r=R$, so we must instead truncate the model at some radius $r_0 > 0$, where we imagine that a stellar bulge or some other non-disky structure forms. In that case we require that the torque have a small value at $r=r_0$, so that the inner bulge region does not do work on the disk.

The evolution of the system is now fully specified. At any given time, equation (\ref{torquedim}) specifies the viscous torque. That torque in turn sets the time evolution of $\Sigma$, $\sigma$, and $\Sigma_*$ via equations (\ref{masscons}), (\ref{veldisp}), and (\ref{sfr}). Before moving on, however, we pause to point out some of the important physical properties embodied in equation (\ref{torquedim}). First, the star formation rate does not appear explicitly in the torque equation. This is because for $\sigma_* \approx \sigma$ changing gas into stars does not significantly affect the stability of the disk, and with our approximate form for $Q$ it does not affect the stability at all. Star formation enters the problem solely through its effects on $f_g$. Second, if $\eta=0$ then clearly $\calT = 0$ is a solution for $\dot{M}_{\rm ext} = 0$. Physically, this represents the fact that, if there is neither dissipation of turbulence nor accretion, then the disk can remain marginally stable without any mass transport.

\subsection{Non-Dimensional Equations}

It is helpful at this point to non-dimensionalize our equations and derive some characteristic numbers. If we make a change of variables $r = xR$, $\sigma = s v_\phi(R)$, $v_\phi= u v_\phi(R)$, and $\calT = \tau \dot{M}_{\rm ext} v_\phi(R) R$, then we can rewrite equation (\ref{torquedim}) as
\begin{equation}
\label{torquenondim}
\tau'' + h_1 \tau' + h_0 \tau = H
\end{equation}
with
\begin{eqnarray}
h_0 & = & \left(\frac{\beta^2 -1}{3f_g-1}\right) \frac{u^2}{x^2 s^2} \\
h_1 & = & -\frac{5(\beta+1) x s'+(3f_g-1)s(\beta+\beta^2+x\beta')}{(3f_g-1)(\beta+1) s x} \\
H & = & \frac{\eta}{\chi} \left(\frac{2 f_g \sqrt{2(\beta+1)^3}}{3f_g-1}\right) \frac{s u^3}{x},
\end{eqnarray}
subject to the boundary condition $\tau' = -\beta-1$ at $x=1$. Here primes indicate differentiation with respect to $x$, and we have defined
\begin{equation}
\chi = \frac{G \dot{M}_{\rm ext}}{v_\phi(R)^3}.
\end{equation}
The form of equation (\ref{torquenondim}) is instructive. The coefficients $h_0$ and $h_1$ appearing on the left-hand side depend on the current state of the disk without reference to external accretion or turbulent dissipation. Those affect only the inhomogeneous term $H$ on the right-hand side, which is proportional to $\eta/\chi$. We may view $H$ as the driving term for the system, with more rapid dissipation of turbulence (i.e.\ larger $\eta$) and lower accretion rates (lower $\chi$) both tending to produce larger torques.

We can similarly non-dimensionalize the evolution equation for the gas surface density and velocity dispersion by defining $\Sigma = S \dot{M}_{\rm ext}/(v_\phi(R) R)$ and $t = T [2\pi R/v_\phi(R)]$, so that the evolution equations become
\begin{eqnarray}
\label{massevolnondim}
\frac{\partial S}{\partial T} & = & \frac{(\beta^2+\beta+x \beta')\tau'-x(\beta+1)\tau''}{(\beta+1)^2 u x^2} -\frac{dS_*}{dT} \\
\label{sevol}
\frac{\partial s}{\partial T} & = & \frac{1}{3 (\beta+1)^2 s S u x^3}
\left\{
u^2 (\beta+1)^2(\beta-1) \tau
\right.
\nonumber \\
& &
\left. {}
+ s x [s(\beta+\beta^2+x\beta')-5 (\beta+1) x s'] \tau'
\right.
\nonumber \\
& &
\left. {}
- (\beta+1) s^2 x^2 \tau'' 
- 2 \pi (\beta+1)^2 \eta s^2 S u^2 x^2
\right\},
\end{eqnarray}
where
\begin{equation}
\frac{dS_*}{dT} = 2\pi \epsilon_{\rm ff} \sqrt\frac{u S^3 \chi}{s x},
\end{equation}
we have defined $\Sigma_* = S_* \dot{M}_{\rm ext}/(v_\phi(R) R)$ in analogy with $S$, and the $Q=1$ condition in dimensionless form is
\begin{equation}
\label{qnondim}
\frac{\sqrt{2(\beta+1)} u s}{\pi \chi x (S+S_*)} = 1.
\end{equation}
In these units the orbital period is $1$, and the fraction of the disk mass that accretion adds per orbit is $\langle S + S_*\rangle^{-1}$, where the angle brackets indicate an average over the disk area. 

\section{Steady-State Disks}
\label{sec:steady}

\subsection{The Steady-State Solution}

Having derived the basic equations that govern the system, we now search for steady-state solutions, which we can obtain analytically. A true steady state is obviously not possible in a real disk that undergoes mass accretion and star formation, but we can find solutions which are steady for time periods that are short compared to the star formation and accretion timescales. We will therefore set $\epsilon_{\rm ff} = 0$, and in Section \ref{steadycondition} we will check that this is a reasonable approximation. In this case a steady state solution is one for which $\partial\dot{M}/\partial r = 0$, or in dimensionless form
\begin{equation}
\frac{d}{dx} \left[\frac{\tau'}{(\beta+1)u}\right] = 0.
\end{equation}
Combined with the boundary condition that $\tau' = -(\beta+1)$ at $x=1$, this implies that the steady state solution is
$\tau' = -(\beta+1) u$. One can immediately verify using equation (\ref{massevolnondim}) that such a torque gives $\partial S/\partial T = 0$.

To make further progress toward an analytic solution, we concentrate our attention on cases where $\beta$ has a constant value. This is a reasonable limitation, since most galaxies have flat rotation curves ($\beta = 0$), while Keplerian disks have $\beta = -1/2$. For constant $\beta$, we have $u = x^\beta$, and we can analytically integrate the steady solution for $\tau'$ to obtain $\tau = -x^{\beta+1} + c$, where $c$ is a constant to be determined by the regularity condition at the inner boundary. In Appendix \ref{innerbc} we provide this analysis for two of the most physically important cases, $\beta = 0$ (flat rotation) and $\beta=-1/2$ (Keplerian rotation), which shows that $c = 0$ in both those cases. Of course $\beta$ cannot have a constant value of $0$ or $-1/2$ all the way to $r=0$, since the rotation velocity would diverge, but our approximation is appropriate in cases where $\beta$ has a constant value over a large dynamic range in radius, and changes only at small radii where a bulge forms (in the case of a flat rotation-curve galaxy) or a boundary layer joins a disk to a star (for a Keplerian disk).

Given the value that $\tau$ must have in order to produce a steady state, we can plug it into the torque equation to determine the corresponding disk properties that are required. For a flat rotation curve, $\beta = 0$, we have
\begin{eqnarray}
\lefteqn{\tau'' - \left(\frac{5}{3 f_g - 1}\right) \frac{s'}{s} \tau' - \frac{1}{(3f_g-1) s^2 x^2} \tau}\qquad\qquad\qquad\qquad\qquad
\nonumber \\
& = & \left(\frac{2^{3/2}f_g }{3 f_g-1}\right) \left(\frac{\eta}{\chi}\right) \frac{s}{x}.
\end{eqnarray}
and the steady-state condition $\tau = -x$ then implies
\begin{equation}
s' = \frac{2\sqrt{2} f_g \eta s^3 - \chi}{5 s x \chi}
\end{equation}
Clearly
\begin{equation}
\label{sigmaeq}
s = \frac{1}{\sqrt{2}}\left(\frac{\chi}{\eta f_g}\right)^{1/3} \quad\mbox{ or }\quad
\sigma = \frac{1}{\sqrt{2}}\left(\frac{G\dot{M}_{\rm ext}}{\eta f_g}\right)^{1/3}
\end{equation}
is an exact solution, and numerical integration shows that all solutions converge to this value very quickly at $x<1$ regardless of the value of $s$ at $x=1$. The corresponding surface density and inward velocity of the material are, from equations (\ref{mdotdefn}) and (\ref{qapprox}),
\begin{eqnarray}
\label{coldeneq}
\Sigma & = & \frac{v_\phi}{\pi G r} \left(\frac{f_g^2 G \dot{M}_{\rm ext}}{\eta}\right)^{1/3} \\
v_r & = & -\eta \frac{\sigma^2}{v_\phi},
\label{vreq}
\end{eqnarray}
and the corresponding dimensionless viscosity parameter \citet{shakura73} and viscous evolution timescales are
\begin{eqnarray}
\label{alphaeq}
\alpha & = & \frac{G \dot{M}}{3 \sigma^3} = \frac{2\sqrt{2}}{3}\eta f_g \\
\label{tvisc}
t_{\rm visc} & = & \frac{R}{v_r(R)} = \left(\frac{f_g^2}{\eta \chi^2}\right)^{1/3} \frac{t_{\rm orb}}{\pi},
\end{eqnarray}
where $t_{\rm orb} = 2\pi R/v_\phi(R)$ is the outer disk orbital period. 
(We omit a factor of $Q$ in the equation for $\alpha$ because it is set to unity in our model.) For the corresponding Keplerian case ($\beta=-1/2$) it is easy to verify by a similar procedure, and using the analysis of the singular point provided in Appendix \ref{innerbc}, that the steady solution is $\tau=-\sqrt{x}$, $s = [3\chi/(4\eta f_g)]^{1/3}$.

Our result is easy to understand intuitively. For $\sigma \gg c_s$, the rate per unit mass at which the turbulence decays is $\eta \sigma^2 \Omega$, so the turbulent decay time is of order the orbital timescale times $\eta$. To keep the velocity dispersion constant, mass must move inward at a rate such that the decrease in gravitational potential energy balances balances this radiative loss. For a flat rotation curve, the decrease in potential involved in moving from radius $r_0$ to radius $r$ is $v_\phi^2\ln (r/r_0)$, so inward drift at a velocity $v_r$ causes the potential energy per unit mass to decrease at a rate $v_\phi^2 (v_r/r)$. Equating the rates of dissipation and energy increase gives $\eta \sigma^2 \Omega = v_\phi^2 (v_r/r)$, and it follows immediately that $v_r = \eta \sigma^2 / v_\phi$.

It is also worth noting that this result is very similar to that of \citet{gammie01a}, who finds that, in steady state in a disk that cools on a timescale $\tau_c$, gravitationally-induced turbulence produces an effective viscosity $\alpha=[\gamma(\gamma-1) (9/4) \Omega\tau_c]^{-1}$. Our effective ``cooling time" for the supersonic turbulence is $\tau_c = 1/(\eta \Omega)$, and the remainder of our result differs from his only in that we have modeled disks with a stellar component, and that our pressure and internal energy are appropriate for supersonically turbulent motion on scales comparable to the disk scale height, rather than for an adiabatic gas described by a polytropic equation of state. The former introduces a dependence on $f_g$, and the latter produces the slight change in leading coefficient. That \citeauthor{gammie01a}'s result can be obtained on energetic grounds, rather than by computing stresses as in his derivation, has also been pointed out by \citet{rice05} and \citet{rafikov09a}.

\subsection{Conditions for Steady-State}
\label{steadycondition}

In deriving the steady solution we have ignored the change in total disk mass due to accretion. More subtly, our steady solution has constant $\dot{M}$ all the way in to $r=0$, so mass effectively vanishes through the origin. These approximations are only reasonable for time scales over which the total disk mass changes little. We can define the accretion timescale for our steady solution as 
\begin{equation}
t_{\rm acc} = \frac{\int_0^R 2 \pi r (\Sigma+\Sigma_*)\, dr}{\dot{M}_{\rm ext}} = \frac{t_{\rm orb}}{\pi(\eta f_g \chi^2)^{1/3}} = \frac{t_{\rm orb}}{2\pi \eta f_g s^2}.
\end{equation}
Note that $t_{\rm acc}$ is the time required for external accretion to double the disk mass, and is distinct from the viscous accretion time defined by equation (\ref{tvisc}). To avoid confusion we will always refer to that quantity as the viscous timescale, and use the accretion time solely to refer to the mass-doubling time produced by external infall. For our fiducial $\eta=3/2$, we have
\begin{equation}
\label{tacc}
t_{\rm acc} \simeq 1.3 f_g^{-1/3} \chi_{0.1}^{-2/3} t_{\rm orb},
\end{equation}
where $\chi_{0.1} = \chi/0.1$. Similarly, we have neglected star formation, which is only reasonable for time short compared to the star formation timescale. We can define the star formation timescale as the mean ratio of star formation rate to gas surface density in the disk:
\begin{equation}
t_{\rm SF} = \frac{ t_{\rm orb}}{\pi} \int_0^1 (2\pi x) \frac{S}{dS_*/dT}\, dx = \frac{t_{\rm orb}}{(162\pi^2)^{1/4} f_g^{1/2}\epsilon_{\rm ff}}.
\end{equation}
For our fiducial $\epsilon_{\rm ff} = 0.01$, this gives 
\begin{equation}
\label{tsf}
t_{\rm SF} \simeq 16 f_g^{-1/2} t_{\rm orb}.
\end{equation}

In comparison, the characteristic evolution timescale for the disk should be the viscous time defined by equation (\ref{tvisc}), since this is the time required to drain the material in the disk and replace it with newly-accreted material. For our fiducial $\eta=3/2$ and a flat rotation curve, this is
\begin{equation}
t_{\rm visc} = 1.3 f_g^{2/3} \chi_{0.1}^{-2/3} t_{\rm orb}.
\end{equation}
Comparing the viscous time to the timescales relevant for accretion and star formation, we have
\begin{eqnarray}
\frac{t_{\rm visc}}{t_{\rm acc}} & = & f_g \\
\frac{t_{\rm visc}}{t_{\rm SF}} & = & \left(\frac{162}{\pi^2}\right)^{1/4} \frac{f_g^{7/6} \epsilon_{\rm ff}}{\eta^{1/3}\chi^{2/3}} = 0.08 f_g^{7/6} \chi_{0.1}^{-2/3},
\end{eqnarray}
for our fiducial values of $\eta$ and $\epsilon_{\rm ff}$. Our neglect of accretion-induced changes in the disk mass and star formation-induced changes in the gas fraction is reasonable when these two ratios are $\la 1$. Clearly the requirement that $t_{\rm visc} / t_{\rm acc} \la 1$ is always satisfied, although perhaps only marginally if $f_g$ is large. The requirement that $t_{\rm visc} / t_{\rm SF} \la 1$ is satisfied as long as $\chi \ga 10^{-3}$, or longer if the disk is non-star-forming ($\epsilon_{\rm ff} = 0$). Thus we expect that our steady state model is reasonable under these conditions.

\section{Time-Dependent Solutions}
\label{sec:nonsteady}

\begin{figure*}
%\begin{figure}
\plotone{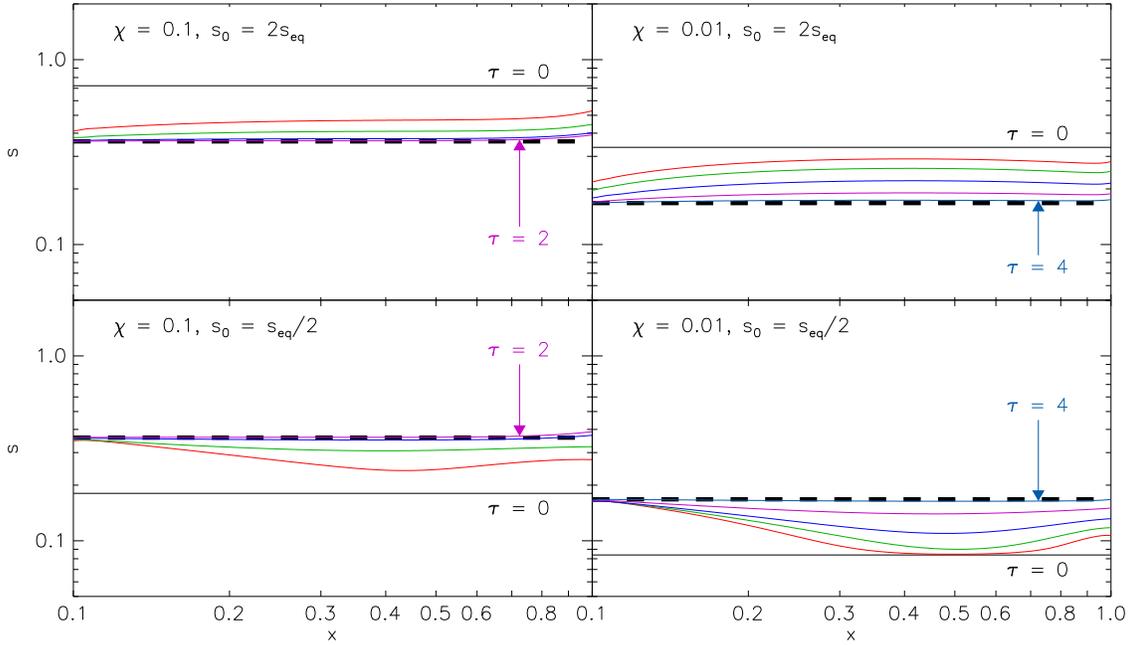}
\caption{
\label{fig:sevol}
Time evolution of the velocity dispersion $s$ as a function of radius $x$ for runs with $\chi=0.1$ (left column) and $\chi=0.01$ (right column), and with initial velocity dispersions $s_0 = 2 s_{\rm eq}$ (top row) and $s_0 = s_{\rm eq}/2$ (bottom row) at all radii. Here $s_{\rm eq} = [\chi/(\eta f_g)]^{1/3}/\sqrt{2}$ (equation \ref{sigmaeq}) is the analytically-computed equilibrium velocity dispersion, and these runs use $\eta=3/2$, $f_g = 1/2$. In each plot, the black curve labeled $\tau=0$ is the initial velocity dispersion, the red curve is the velocity dispersion at $\tau=0.25$ (i.e.\ after $0.25$ outer orbits), and each subsequent curve after that represents a factor of 2 increase in time: $\tau = 0.5$ (green), $\tau = 1$ (blue), $\tau=2$ (purple), and, for the $\chi=0.01$ runs, $\tau=4$ (aqua). The thick dashed black line is $s=s_{\rm eq}$.
\\
}
\end{figure*}
%\end{figure}

We now consider a disk that is initially out of steady state, with a specified initial value of $\chi$ and $f_g$. For the reasons discussed in the previous section, we take $\epsilon_{\rm ff} = 0$ and $\chi$ and $f_g$ as constant. We start each calculation from an initial velocity dispersion specified on a grid of $N_x$ cells, logarithmically spaced. The grid runs from $x=x_0$ to 1, where $x_0$ is close to zero. For the boundary conditions, we cannot simultaneously require that $\tau$ obey the regularity condition derived in Appendix \ref{innerbc} at the inner boundary and that $\tau' = -(\beta+1) u$ at the outer boundary; this would amount to applying three boundary conditions to a second-order ODE, and for general choices of $s$ no solution for $\tau$ exists that satisfies all three conditions. Instead, we continue to fix $\tau'$ (and thus the accretion rate) at the outer boundary, and at the inner boundary we require that the torque be $\tau=-x_0$. This choice amounts to requiring that the work done by gas in the region $x<x_0$ on the computational domain at $x\geq x_0$ approaches $0$ as $x_0\rightarrow 0$, while the mass flux through the inner boundary at $x=x_0$ is allowed to vary freely. This is a good approximation to a disk being fed from the outside a fixed rate and that has an inner bulge region or an inner boundary layer that is stress free, but which can accept mass at varying rates. Note that we do not set $\tau=0$ at $x=x_0$ exactly, because this is inconsistent with the steady state solution.

With this setup, we evolve the system according to equation (\ref{sevol}) using the algorithm given in Appendix \ref{updatealgorithm}\footnote{A Mathematica program to implement this algorithm is available at http://www.ucolick.org/$\sim$krumholz/downloads.html.}. Note that we find that the unmodified evolution equation (\ref{sevol}) for $s$ is numerically unstable to the growth of small oscillations on the grid scale. We damp these by adding a small amount of viscosity to the disk evolution, implemented in a manner that maintains exact energy conservation.

In Figure \ref{fig:sevol} we show the evolution of disks with $\beta=0$ (flat rotation curve), $f_g = 0.5$, $\eta=3/2$, $x_0 = 0.1$ and $N_x = 500$. The left panels show runs with $\chi=0.1$, and the right panels show runs with $\chi = 0.01$. The top row shows disks with an initial velocity dispersion $s_0$ equal to either twice the equilibrium value $s_{\rm eq}$ given by equation (\ref{sigmaeq}), and the bottom row shows disks with an initial velocity dispersion that is half this value. As the figure shows, all disks evolve toward the equilibrium solution found in Section \ref{sec:steady} very rapidly, regardless of whether they start with initial velocity dispersions smaller or larger than $s_{\rm eq}$. The runs with $\chi=0.1$ halve their distance from the equilibrium solution in less than a quarter of an outer orbital period, and converge to within 10\% of the equilibrium solution within one full outer orbit, after which point they are essentially static.\footnote{The $\chi=0.1$ cases miss the equilibrium value very slightly and converge to a velocity dispersion that is a few percent above it. This is an artifact of the small viscosity we require in order to maintain numerical stability in this run. The $\chi=0.01$ cases are stable with a somewhat smaller viscosity, so the deviation from the exact equilibrium is unnoticeable for them.} in fact, the time required to reach equilibrium seems to be a factor of $\sim 2-3$ less than our naive estimate that it should the viscous time $t_{\rm visc}$. This may be because $t_{\rm visc}$ is an estimate of the time for the material to reach zero radius, while in our case the material need not travel as far to set up an equilibrium velocity dispersion and surface density profile.

Thus the time required to reach equilibrium is well below the accretion time $t_{\rm acc} = 1.6 t_{\rm orb}$ (for $f_g=1/2$) we computed in Equation (\ref{tacc}), and is far less than the star formation time $t_{\rm SF} = 20 t_{\rm orb}$ given by Equation (\ref{tsf}), in accord with our analytic expectations. We therefore conclude that disks with $\chi = 0.1$ converge to the equilibrium configuration on a timescale short compared to both the accretion and star formation timescales. For $\chi = 0.01$ the convergence to equilibrium is slightly slower, but the runs are within 10\% of equilibrium by 2 outer orbits, and are within $\sim 1\%$ of equilibrium by 4 outer orbits. Since the accretion timescale is $7.6$ orbital times for $\chi = 0.01$, and the star formation time is $20 t_{\rm orb}$, these runs too reach equilibrium fast compared to $t_{\rm acc}$ or $t_{\rm SF}$.

Thus we have demonstrated that the time-independent solution we obtained in Section \ref{sec:steady} is not only an exact equilibrium, it is an attractor toward which initially out-of-equilbrium disks will converge. As long as $\chi \ga 10^{-3}$ (or for arbitrarily small $\chi$ in non-star-forming disks), this convergence occurs on a timescale short compared to either the either the star formation timescale or the accretion timescale over which the disk mass changes appreciably. This means that our earlier decision to neglect both star formation and changes in the rotation curve due to accretion is reasonable, and that a disk that is out of steady state will converge to its time-independent equilibrium state faster than either star formation or accretion can alter that equilibrium. It also implies a vast simplification in comparing to observations: since convergence to equilibrium is fast, we can generally assume that the velocity dispersions and surface densities of observed disks reflect the instantaneous equilibrium state dictated by their current external accretion rates and gas fractions. Of course we still have not included star formation feedback, a topic we approach in Section \ref{sec:feedback}.

\section{Discussion}
\label{sec:discussion}

\subsection{Cosmological Evolution of the Velocity Dispersion of Galactic Disks}
\label{sec:diskevol}

\begin{figure}
\plotone{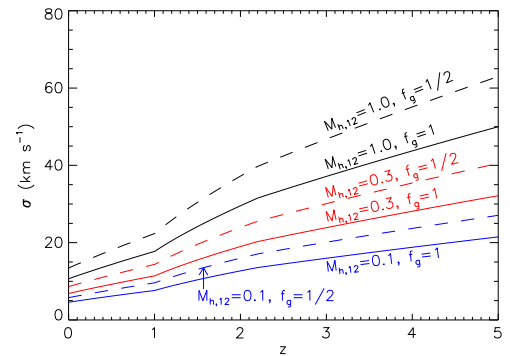}
\caption{
\label{fig:halosigma}
Disk velocity dispersion $\sigma$ versus redshift $z$ for halos of mass $M_{h,12} = 0.1$, $0.3$, and $1.0$ (blue, red, and black lines) and gas fraction $f_g = 1/2$ or $1$ (dashed and solid lines). The plot uses our fiducial $\eta = 3/2$ and assumes $f_{b,0.18} = 1/2$, i.e.\ half the infalling baryons are gas and half are stars. The $f_g = 1$ case is appropriate for systems where there no stars or where $\sigma_* \gg \sigma$, such as present-day galactic disks, while the case $f_g=1/2$ is appropriate for high-redshift disks where $\sigma_* \approx \sigma$.
}
\end{figure}

We have now shown that disks dominated by gravitationally-driven turbulence rapidly converge to an equilibrium state in which their velocity dispersions are determined by their gas fractions and accretion rates. Since this convergence happens on an orbital timescale, most galactic disks should be found near their equilibrium state. We can use this result, coupled to a simple model for how galaxy halos accrete mass, to study the evolution of disk velocity dispersions over cosmic time. Using Press-Schechter fits to dark matter simulations, \citet{neistein08a} estimate the mean dark matter accretion rate onto halos of a given mass at a given redshift. \citet{bouche09a} extend this to give an estimate of the gas accretion rate onto the disk at the center of the halo, which we adopt:
\begin{equation}
\label{mdotgas}
\dot{M}_g = 7.0\epsilon_{\rm in} f_{b,0.18} M_{h,12}^{1.1} (1+z)^{2.2}\, \msun\mbox{ yr}^{-1},
\end{equation}
where $z$ is the redshift, $f_{b,0.18}$ is the gas mass fraction of the infall divided by 0.18, the universal baryon fraction, $M_{h,12}$ is the halo mass in units of $10^{12}$ $\msun$, and $\epsilon_{\rm in}$ is the fraction of gas entering the halo that reaches the galactic disk rather than being shock-heated and joining the halo. This is approximately given by
\begin{equation}
\epsilon_{\rm in} = \left\{
\begin{array}{ll}
0.7 f(z), \quad & M_{h,12} < 1.5 \\
0, & M_{h,12} > 1.5
\end{array}
\right.,
\end{equation}
where $f(z)$ is a function that is linear in time and varies from unity at $z=2.2$ to 0.5 at $z=1$.\footnote{We compute the time as a function of redshift, and all other cosmology-dependent quantities, using $\Omega_m = 0.28$, $\Omega_{\Lambda} = 0.72$, and $h=0.70$.} Inserting $\dot{M}_g$ from equation (\ref{mdotgas}) for $\dot{M}_{\rm ext}$ in equation (\ref{sigmaeq}), we are able to evaluate the expected velocity dispersion of gravitational instability-dominated galactic disks as a function of halo mass and redshift. We do so in Figure \ref{fig:halosigma}.

Examining the plot, we see that for a Milky Way-like halo ($M_{h,12} = 1$, \citealt{xue08a}) where $\sigma_* \gg \sigma$ (so that the $f_g=1$ case applies), we predict a typical velocity dispersion of $10.7$ km s$^{-1}$. While this is very slightly higher than the value of $\sigma \simeq 8$ km s$^{-1}$ observed in typical Milky Way-like disks today (\citealt{blitz04}, \citealt{dib06a}, and references therein), the agreement is quite good given our purely analytic model. Our results are quite similar to the numerical ones obtained by \citet{kim07} and \citet{agertz09a}. Moreover, as \citeauthor{agertz09a}\ point out, gravitationally-driven turbulence has the advantage that it can operate even in the outer H~\textsc{i} disk where there is very little star formation, so mechanisms such as supernovae that are invoked to explain turbulence in the inner disk \citep[e.g.][]{de-avillez07a, joung09a} are unavailable. We also predict lower velocity dispersion in smaller halos, and this appears to be consistent with the somewhat lower H~\textsc{i} velocity dispersions seen in dwarf galaxies \citep{walter08a, chung09a}. Finally, however, we do note that there are alternative models to explain outer disk turbulence, including magnetorotational instability \citet{sellwood99a, piontek07a} and accretion of clumpy gas \citet{santillan07a}.

Within the same framework we are able to explain the large velocity dispersions of $20-80$ km s$^{-1}$ found in galactic disks found at redshifts $\sim 1.5-3$ \citep{cresci09a}. The observed galaxies likely correspond to $\sim 10^{12}$ $\msun$ halos. For redshifts in this range and $f_g \sim 1/2$, typical of galaxies at that redshift \citep{daddi09a, tacconi10a} we predict typical velocity dispersions of $30-50$ km s$^{-1}$, with fluctuations at the factor of $\sim 1.5$ level, corresponding to the expected factor of $\sim 3$ level variations in the accretion rates of halos at the same mass and redshift. This is in good agreement with the observations.

\begin{figure}
\plotone{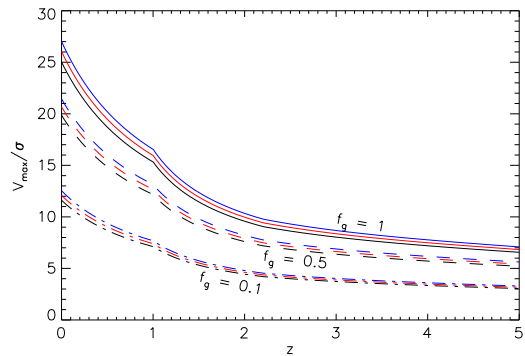}
\caption{
\label{fig:halovsigma}
Ratio of disk maximum circular velocity $V_{\rm max}$ to velocity dispersion $\sigma$ as a function of redshift $z$ for halos of mass $M_{h,12} = 0.1$, $0.3$, and $1.0$ (blue, red, and black lines, top to bottom) and gas fraction $f_g = 0.1$, $0.5$, or $1$ (dot-dashed, dashed, and solid lines). All parameters are the same as for Figure \ref{fig:halosigma}.
}
\end{figure}

It is also instructive to compare the velocity dispersions we predict to the expected rotation velocities of galactic disks. We compute the approximate virial velocity of a halo as a function of mass and redshift following the approximation given in Appendix A2 of \citet{dekel06a}, and we take the maximum circular velocity to be $1.2$ times this based on fitting the zero point of the Tully-Fisher relation \citep{dutton07a}. With this approximation, we plot $V_{\rm max}/\sigma$ in Figure \ref{fig:halovsigma}. We see that accretion-driven turbulence naturally produces the transition from disks with $V_{\rm max}/\sigma\sim 5$ found at redshifts $\ga 2$ to disks with $V_{\rm max}/\sigma \sim 20-25$ found today. 

Interestingly, we find that there is little dependence of $V_{\rm max}/\sigma$ on halo mass. Instead, the primary dependence is in $f_g$, the gas mass fraction; analytically, $V_{\rm max}/\sigma \propto f_g^{-1/3}$. Thus the most gas-dominated systems (or old galaxies that have $\sigma_* \gg \sigma$) have the largest $V_{\rm max}/\sigma$, while gas-poor systems have smaller $V_{\rm max}/\sigma$. This suggests that the range of $V_{\rm max}/\sigma$ seen for galaxies at $z\sim 2$ by the SINS survey \citep{cresci09a} represents a sequence in gas fraction. The dispersion-dominated galaxies should on average be comparatively gas poor, while rotation-dominated ones should be gas rich. Of course fluctuations in accretion rate can also cause changes in $\sigma$, so detecting this effect will require samples large enough for this noise source to be averaged out. Nonetheless, it seems likely that data to test this prediction will become available in the next few years.

\subsection{High-Redshift Galaxies}
\label{sec:highz}

It is particularly interesting to apply our models to the $z\sim 2-3$ galaxies observed by \citet{elmegreen04b, elmegreen05a}, \citet{genzel08a}, \citet{forster-schreiber09a}, \citet{cresci09a}, and others, since these are thought to be examples of strongly gravitational instability-dominated disks. We first note that, in the redshift range $z=2-3$ for halos of mass $M_h = 10^{12}$ $\msun$, thought to be typical of the observed systems, the models shown in Figures \ref{fig:halosigma} and \ref{fig:halovsigma} give $\chi = 8\times 10^{-3} - 1.1\times 10^{-2}$. For these values of $\chi$ and gas fractions $f_g = 1/2$, using equations (\ref{tacc}) and (\ref{tsf}) the ratio of star formation time to accretion time $t_{\rm SF} / t_{\rm acc} = 1.8 - 2.6$, so the star formation rate is roughly $1/2 - 1/3$ of the total accretion rate. Given the uncertainties in this model and the dispersion in expected accretion rates, for simplicity we can simply adopt $\dot{M}_* \approx \dot{M}_{\rm ext}$. Since the star formation rates are observed (and have typical values $\sim 100$ $\msun$ yr$^{-1}$), we can plug them into our model in place of $\dot{M}_{\rm ext}$ in order to predict disk properties.

Doing so, we find that the redshift $2-3$ disks should have velocity dispersions (from equation \ref{sigmaeq})
\begin{equation}
\sigma \approx 47\mbox{ km s}^{-1}\, f_g^{-1/3} \dot{M}_{*,100}^{1/3},
\end{equation}
where $\dot{M}_{*,100} = \dot{M}_* / 100$ $\msun$ yr$^{-1}$, independent of their maximum rotation velocities $V_{\rm max}$. Thus galaxies of similar star formation rate and gas fraction should have the same $\sigma$ independent of $V_{\rm max}$. The viscous accretion timescale required for the gas at the edge of one of these disks to reach the center is (from equation \ref{tvisc})
\begin{equation}
t_{\rm visc} \approx 600\mbox{ Myr } f_g^{2/3} R_{10} V_{200}^{-1} \dot{M}_{*,100}^{-2/3},
\end{equation}
where $R_{10} = R/10$ kpc and $V_{200} = V_{\rm max}/200$ km s$^{-1}$, and the gas mass is (from equation \ref{coldeneq})
\begin{eqnarray}
\nonumber
M_g & = & R v_{\phi} \left(\frac{f_g^2 \dot{M}_*}{\eta}\right)^{1/3} \\
& = & 3\times 10^{10}\,\msun\; f_g^{2/3} R_{10} V_{200} \dot{M}_{*,100}^{1/3}.
\end{eqnarray}
The ratio of baryonic to dynamical mass within the disk region $R$ is
\begin{equation}
\frac{M_{\rm bar}}{M_{\rm dyn}} = 0.3\, f_g^{-1/3} V_{200}^{-1} \dot{M}_{*,100}^{1/2}.
\end{equation}
To the extent that these quantities have been observed, they are in good agreement with the results of our model.

It is also useful to verify that our approximation that $\sigma_* \approx \sigma$ is valid for these galaxies. Once stars form, transient spiral structures will dynamically heat them until the stellar disk becomes stable against further spiral patterns \citep{sellwood84a, carlberg85a}. The characteristic timescale for this heating is $[(Q_{\rm lim} - Q_*)/\tau] t_{\rm orb}$, where $Q_{\rm lim}\approx 2$ is the limiting value at which the disk becomes stable against spiral perturbations, $Q_*$ is the current Toomre $Q$ parameter for the stars, and numerical experiments show that $\tau\sim 4-5$ for $t_{\rm orb}$ evaluated at one disk scale length. If we assume that the stars are born at $Q_* = 1$, then the characteristic timescale over which the heat is
\begin{equation}
t_{\rm heat} \approx 750\mbox{ Myr } R_{10} V_{200}^{-1},
\end{equation}
where we have taken the radial scale length to be half of $R$. In contrast, the time required to double the stellar mass is
\begin{equation}
t_{\rm *} = \frac{1-f_g}{f_g}\left(\frac{M_g}{\dot{M}_*}\right) = 300\mbox{ Myr } \frac{1-f_g}{f_g^{1/3}} R_{10} V_{200} \dot{M}_*^{-2/3}.
\end{equation}
Thus we see that the time required for stars to increase their velocity dispersion via spiral structure is generally comparable to or longer than the time required for a new generation of stars to form with the same velocity dispersion as the gas. Our approximation that the stellar population has the same velocity dispersion as the gas in these galaxies is therefore reasonable.

\subsection{Effects of Stellar Feedback}
\label{sec:feedback}

In our idealized models, we have neglected the influence of stellar feedback by setting $\calG = 0$. This is obviously reasonable if we are concerned with the outer parts of a galactic disk where there is no star formation, or a protostellar disk where at most a few stars will form. It is not reasonable for the centers of present-day galactic disks, where supernovae are clearly important. It is questionable whether stellar feedback is important in ULIRGs or the high surface density galaxies found in the early universe. Supernovae are not effective in such environments \citep{thompson05, joung09a}, but stellar radiation pressure may be. Whether it can actually drive the observed velocity dispersions in high redshift galaxies is a matter of debate \citep{murray09a, krumholz09d}.

In those situations where feedback is significant, we can qualitatively see how it would change our results by noting that adding a non-zero $\calG$ to our equations would have roughly the same effect as lowering $\eta$. Physically, if star formation injects turbulence into the ISM at an appreciable rate, this is equivalent to reducing the rate at which turbulence decays -- we effectively increase the ``cooling time" of the disk. Consulting equations (\ref{sigmaeq}) -- (\ref{alphaeq}), we see that the effect of this is to increase the velocity dispersion and surface density in the equilibrium state, while reducing the radial velocity and the rate of angular momentum transport.

We caution that this analysis is only valid as long as the feedback is not too strong. In particular, we require that $\calL$ remains larger than $\calG$ for a $Q=1$ disk, and that the turbulent stresses created by the feedback mechanism are significantly weaker than the stresses induced by gravitational instability-driven turbulence. If the first requirement is not met, then feedback will drive the velocity dispersion up to the point where $Q>1$, and the gravitational instability will shut off. If the latter condition fails, then gravitational instability will continue, but our calculation of the transport rate will not be correct because we have not included stresses induced by feedback. Even if both requirements are met, our analysis of feedback effects should be regarded as qualitative rather than quantitative. Energy injection $\calG$ appears on the right hand side of the torque equation with the opposite sign as $\calL$, but their functional dependence on other disk parameters (surface density, velocity dispersion, etc.) are almost certainly different. The exact effects of feedback will depend on how energy injection varies with these quantities, which will in turn depend on the type of feedback and the physics of the ISM.

\subsection{Protostellar Disks}
\label{sec:protostellar}

Although we have focused our discussion thus far on galactic disks, our model applies for arbitrary rotation curves, gas fractions, and infall rates, so we can apply it equally well to protostellar disks. To understand the expected levels of turbulence in protostellar disks, we use the parameterization of infall due to \citet{kratter08a, kratter10a}, who introduce the dimensionless numbers:
\begin{equation}
\xi = \frac{G \dot{M}_{\rm ext}}{c_{s,d}^3}
\qquad\qquad
\Gamma = \frac{\dot{M}_{\rm ext}}{M_{*d} \Omega_{k,\rm in}},
\end{equation}
where $c_{s,d}$ is the sound speed in the disk, $M_{*d}$ is the total mass of the disk and star and $\Omega_{k,\rm in}$ is the Keplerian orbital period of the infalling material. Physically, $\xi$ represents the ratio of the external accretion rate to the maximum rate ($\sim c_{s,d}^3/G$) at which a stable disk can process material, while $\Gamma$ represents (up to a factor of $2\pi$) the fraction by which the disk plus star mass changes per outer disk orbit. Indeed, since $v_{\phi}(R) = R\Omega_{k,\rm in} = \sqrt{G M_{*d}/R}$, with a little algebra it is easy to show that, in the case of a Keplerian disk consisting entirely of gas, our $\chi$ simply reduces to \citeauthor{kratter10a}'s $\Gamma$ parameter.

With this understanding, we can explain the observation by \citet{kratter10a} that, in their simulations, the typical velocity dispersion of disks that do not fragment is comparable to the disk thermal sound speed (see their Figure 8). For a purely gaseous Keplerian disk, our model gives $s = [3\chi/(4\eta)]^{1/3}$, and \citeauthor{kratter10a} show that the disk sound speed is related to the Keplerian velocity at the disk edge by $c_{s,d}/v_\phi(R) = (\Gamma/\xi)^{1/3}$ (their Equation 18). Combining these two results, the expected Mach number of the accretion-driven turbulence is
\begin{equation}
\label{machdisk}
\mathcal{M} = \frac{\sigma}{c_{s,d}} = s \frac{c_{s,d}}{v_\phi(R)} = \left(\frac{3\xi}{4\eta}\right)^{1/3}.
\end{equation}
Since fragmentation is avoided only for disks with $\xi$ of no more than a few, we can take $\xi \sim 1$, and it immediately follows that the expected Mach number $\mathcal{M} \sim 1$. 

We can apply a similar analysis to real protostellar disks: the Mach number of the turbulence in these disks should follow equation (\ref{machdisk}). This means that disks accreting with $\xi \sim 1$, corresponding to $\dot{M}_{\rm ext} \sim 10^{-5}$ $\msun$ yr$^{-1}$ for typical out disk temperatures $T\sim 50$ K, should have disks whose turbulent velocity dispersions are roughly transonic. This state should prevail during the majority of the main accretion phase. Once the main accretion phase ends and the accretion rate drops, the turbulent velocity dispersion should drop to subsonic values. This represents another prediction from our analysis: class 0 and class I protostars should have disks with transsonic turbulent velocity dispersions, while class II and class III sources should have subsonic turbulent velocity dispersions. As ALMA comes online in the next few years and provides resolved molecular line maps of protostellar disks at a variety of stages in their evolution \citep[e.g.][]{krumholz07d}, we will be able to test this prediction.

\subsection{On the Validity of a Local Viscous Approximation for Gravitational Instability-Induced Transport}
\label{locality}

The central approximation we make in our model is that transport of mass, angular momentum, and energy produced by gravitationally-driven turbulence can be represented with a local viscous stress tensor. The validity of this approximation has been the subject of great debate in the past decade. \citet{balbus99a} show that self-gravitating disks cannot in general be modeled with a viscous formalism, but that such an approximation may be reasonable for disks near $Q=1$, the condition that we adopt throughout this work, and that appears to apply to the galactic and protostellar disks we are interested in studying. Based on a combination of analytic arguments and local simulations, \citet{gammie01a} argues that a local prescription is applicable to $Q=1$ disks that are sufficiently thin, $s \la 0.12$, and more recent global simulations \citep{lodato04a, lodato05a, boley06a, cossins09a} generally support this result. \citeauthor{gammie01a}'s condition for a local transport approximation to apply is well-satisfied for galactic disks at redshifts $z \la 2$ (\S~\ref{sec:diskevol}) and for non-fragmenting protostellar disks (\S~\ref{sec:protostellar}). It is marginally violated for the observed disks at $z\sim 2$ (\S~\ref{sec:highz}), suggesting that our model should be considered with some caution for them. At a minimum the thickness of these disks likely produces different fragmentation behavior than a standard thin disk analysis would suggest \citep{begelman09a}.

\section{Summary}
\label{sec:conclusion}

In this paper we derive the basic evolution equations for a disk of gas and stars kept in a state of marginal gravitational instability by a combination of external accretion, inward migration of gas, and decay of turbulent motions due to radiative shocks. In such a disk, we use the equations of conservation of mass, angular momentum, and energy to derive an equation (\ref{torquenondim}) that characterizes the instantaneous rates of mass and angular momentum transport required to maintain the state of marginal stability, and we show that this equation has an analytic steady-state solution in which the disk velocity dispersion (Equation \ref{sigmaeq}), surface density (Equation \ref{coldeneq}), and rates of transport (Equations \ref{vreq} and \ref{alphaeq}) through the disk are determined by the rate of external infall onto the disk and the gas mass fraction within it. We show that disks converge to this steady state on timescales of order the orbital time, much less than the time over which either the rotation curve or the gas mass fraction changes significantly.

Based on our analytic solution for the properties of a gravitational instability-dominated disk and their dependence on the gas mass fraction and the infall rate, we are able to gain new insight into several processes. We show that the velocity dispersions of both the outer H~\textsc{i} disks of present day galaxies and the main disks of redshift $\sim 2$ galaxies can be understood naturally if they are in a state of gravitational instability-regulated equilibrium. Moreover, we can understand the general progression of galactic disks from low values of rotation speed to velocity dispersion ratio, $V_{\rm max}/\sigma$, at high redshift to much higher values today. This progression is driven primarily by a falloff in galaxy accretion rates and secondarily by the development of disks with stellar velocity dispersion much lager than the gas velocity dispersion, reducing the importance of stars in setting the gravitational instability condition. We also predict that the observed range of $V_{\rm max}/\sigma$ values seen at $z\sim 2$ is primarily a sequence in gas mass fraction. Finally, we use the same model to study the velocity dispersions of protostellar disks. We show that our results are in good agreement with numerical simulations of gravitational instability in disks, and we predict that velocity dispersions should be transsonic in class 0 and I protostars, dropping to subsonic for class II and III sources.

Although our attention in this paper is focused on cases that can be solved analytically or nearly so, we close by pointing out that our model, as a result of its grounding in the basic equations of fluid dynamics, is also amenable to a more general numerical treatment. One can easily relax our assumptions of constant gas fraction, negligible influence from stellar feedback, and a fixed relationship between gas and star velocity dispersion. The resulting equations are identical to the ones we have already solved, except that they would need to be solved numerically. There is no fundamental barrier to doing so however, and the result would be a new method for simulating the evolution of marginally unstable star-forming disks that is intermediate between purely analytic models such as those we have pursued here and full numerical simulations that can be extremely costly. We plan to pursue this avenue in future work.

\acknowledgements We thank A.\ Dekel, P.\ Garaud, R.~S.\ Klessen, D.~N.~C.\ Lin, and R.\ Murray-Clay for helpful conversations, G.\ Bertin, A.\ Dekel, J.\ Forbes, K.\ Kratter, G.\ Lodato, C.\ McNally, and K.\ Rice for comments on the manuscript, and the anonymous referee for a helpful report. AB thanks his colleagues at the astronomy department at UCSC for their hospitality and support. Financial support for this work was provided by: an Alfred P.\ Sloan Fellowship (MRK); NASA through ATFP grant NNX09AK31G (MRK); NASA as part of the Spitzer Theoretical Research Program, through a contract issued by the JPL (MRK); the National Science Foundation through grant AST-0807739 (MRK); a Max-Planck-Fellowship and the DFG Cluster of Excellence ``Origin and Structure of the Universe" (AB).

\begin{appendix}

\section{Derivation of the Transport Equations}
\label{derivation}

Here we derive the evolution equations for a thin, axisymmetric disk evolving following the general fluid equations (\ref{continuity}), (\ref{navierstokes}) and (\ref{firstlaw}) including star formation. The derivation follows the same general outline as the standard treatment of disks \citep[e.g.][]{shu92, balbus98a}, with additional terms added to describe star formation and some subtleties that arise in how to treat the energy content of supersonic turbulence. We treat these following the method of \citet{krumholz06d}

Writing out equation (\ref{continuity}) in cylindrical coordinates chosen so that the disk lies in the $z=0$ plane, dropping terms that are zero in axisymmetry, and integrating over $z$ gives equation (\ref{eq:jcons}), which we repeat here for convenience:
\begin{equation}
\label{continuity2d}
\frac{\partial}{\partial t} \Sigma = -\frac{1}{r}\frac{\partial}{\partial r}(r\Sigma v_r) - \dot{\Sigma}_* = \frac{1}{2\pi r}\frac{\partial}{\partial r}\dot{M} -\dot{\Sigma}_*,
\end{equation}
where $\Sigma=\int \rho \, dz$ is the gas surface density, $\dot{\Sigma}_*=\int \dot{\rho}_*\,dz$ is the star formation rate per unit area, $v_r$ is the radial component of the velocity, and we have defined
\begin{equation}
\label{mdotdefn}
\dot{M} \equiv -2\pi r \Sigma v_r
\end{equation}
as the inward radial mass flux.

Writing out the $\phi$ component of the Navier-Stokes equation (\ref{navierstokes}) and performing a similar integration over $z$ yields
\begin{equation}
\Sigma\left[\frac{\partial}{\partial t} v_\phi + \frac{v_r}{r}\frac{\partial}{\partial r}(rv_\phi)\right] = \int \frac{1}{r^2} \frac{\partial}{\partial r} (r^2 T_{r\phi})\,dz
\end{equation}
where $v_\phi$ is the $\phi$ component of the velocity and $T_{r\phi}$ is the $r\phi$ component of the pressure tensor. Multiplying this equation by $2\pi r^2$ gives the evolution equation for the angular momentum (\ref{eq:jcons}):
\begin{equation}
\label{angmom1}
2\pi r \Sigma\left(\frac{\partial}{\partial t} j + v_r\frac{\partial}{\partial r}j\right) = \frac{\partial}{\partial r}  \int 2\pi r^2 T_{r\phi}\,dz  = \frac{\partial}{\partial r} \calT
\end{equation}
where $j=rv_\phi$ is the specific angular momentum of the gas and $\calT = \int 2 \pi r^2 T_{r\phi}\, dz$ is the viscous torque on the gas.

The gas velocity dispersion is determined by energy conservation. Taking the dot product of $\vecv$ with the Navier-Stokes equation (\ref{navierstokes}) and using the continuity equation (\ref{continuity}) to re-arrange, yields
\begin{equation}
\frac{\partial}{\partial t} \left(\frac{1}{2}\rho v^2\right) + \nabla \cdot\left(\frac{1}{2}\rho \vecv v^2\right) =
-\vecv\cdot\nabla p - \rho \vecv\cdot \nabla \psi + \vecv\cdot\nabla\cdot \vecT - \frac{1}{2}\dot{\rho}_* v^2.
\label{energy1}
\end{equation}
Using the continuity equation (\ref{continuity}), we can rewrite the gravitational work term as
\begin{eqnarray}
-\rho \vecv \cdot \nabla \psi & = & -\nabla\cdot (\rho \vecv\psi) + \psi \nabla\cdot (\rho \vecv) \\
& = & -\nabla\cdot (\rho \vecv\psi) - \frac{\partial}{\partial t}(\rho \psi) + \rho\frac{\partial \psi}{\partial t} - \dot{\rho}_*\psi.
\end{eqnarray}
Substituting this into equation (\ref{energy1}) gives the evolution equation for the non-thermal energy:
\begin{equation}
\frac{\partial}{\partial t} \rho \left(\frac{v^2}{2} + \psi \right) + \nabla \cdot\rho \vecv \left(\frac{v^2}{2}+\psi\right) =
-\vecv\cdot\nabla p + \rho \frac{\partial\psi}{\partial t} + \vecv\cdot\nabla\cdot \vecT  - \dot{\rho}_* \left(\frac{v^2}{2}+\psi\right).
\label{nonthermal}
\end{equation}

To include the internal energy, we make use of the first law of thermodynamics, equation (\ref{firstlaw}). Combining this with the continuity equation yields
\begin{equation}
\label{intenergy}
\frac{\partial}{\partial t}(\rho e) + \nabla \cdot \rho \vecv\left(e+\frac{p}{\rho}\right) = \vecv\cdot \nabla p -\dot{\rho}_* e + \Phi + \Gamma - \Lambda.
\end{equation}
Adding equations (\ref{nonthermal}) and (\ref{intenergy}) yields the total energy equation:
\begin{equation}
\frac{\partial}{\partial t} \rho \left(\frac{v^2}{2} + e + \psi \right) + \nabla \cdot\rho \vecv \left(\frac{v^2}{2}+e+\psi+\frac{p}{\rho} \right) = \rho \frac{\partial\psi}{\partial t} + 
\vecv\cdot\nabla\cdot\vecT + \Phi - \dot{\rho}_* \left(\frac{v^2}{2} + e + \psi\right) 
 + \Gamma - \Lambda.
\label{totenergy1}
\end{equation}

We now integrate over $z$ and use our axisymmetric thin disk assumption to drop all terms involving either $z$ velocities or $\phi$ derivatives. This gives
\begin{equation}
\frac{\partial}{\partial t} \Sigma \left(\frac{v^2}{2} + e + \psi \right) + \frac{1}{r}\frac{\partial}{\partial r} \left[r \Sigma v_r \left(\frac{v^2}{2}+e+\psi+\frac{p}{\rho} \right)\right]
= \Sigma \frac{\partial\psi}{\partial t} + \frac{1}{2\pi r}\frac{\partial}{\partial r}(\Omega \calT) 
- \dot{\Sigma}_* \left(\frac{v^2}{2} + e + \psi\right) 
+ \calG - \calL.
\label{totenergy2}
\end{equation}
where $\Omega=v_{\phi}/r$,  $\calG = \int \Gamma\, dz$, and $\calL = \int \Lambda \, dz$. In deriving this equation we have assumed $\rho=\Sigma \delta(z)$ and $\dot{\rho}_* = \dot{\Sigma}_* \delta(z)$, so in this and all subsequent equations we understand that all the terms in parentheses are to be evaluated in the plane $z=0$.

It is convenient to rewrite the kinetic energy plus thermal energy term $v^2/2 + e$ as a sum of terms representing bulk motions on length scales comparable to the radial extent of the galactic disk, small-scale turbulent motions on scales comparable to the disk scale height, and true thermal energy:
\begin{equation}
\frac{1}{2}v^2 + e = \frac{1}{2}\left(v_r^2 + v_\phi^2\right) + \frac{3}{2}\left(\sigma_{\rm nt}^2 +\sigma_{\rm t}^2\right) = \frac{1}{2}\left(v_r^2 + v_\phi^2\right) + \frac{3}{2}\sigma^2
\end{equation}
Here $\sigma_{\rm nt}$ is the one-dimensional non-thermal velocity dispersion of the turbulent motions on the size scale of the disk scale height, $\sigma_{\rm t}^2=(2/3) e$ is the thermal velocity dispersion, and $\sigma^2 = \sigma_{\rm t}^2 + \sigma_{\rm nt}^2$.\footnote{Note that the coefficient $2/3$ in the relation between $\sigma_{\rm t}$ and $e$ is appropriate for a monatomic ideal gas, i.e.\ $\gamma=5/3$; for molecular gas $\gamma$ can have a different value if the temperature is high enough to excite rotational levels of H$_2$ or to induce changes in the ortho- to para-H$_2$ ratio. However, for galaxies with a molecule-dominated ISM, $\sigma_{\rm nt}$ is always small compared to $\sigma_{\rm t}$, so we need not worry about a small variations in the coefficient of $\sigma_{\rm t}$.} It is somewhat less clear how to evaluate the pressure $p$ in terms of $\sigma_{\rm t}$ and $\sigma_{\rm nt}$. The microphysical pressure is $p=\rho \sigma_{\rm t}^2$, but if we are averaging over scales much larger than the characteristic size of the turbulent eddies (which is of order the disk scale height), then the eddies provide an additional effective pressure, and we will instead have $p=\rho \sigma^2$. Since we are interested in the large-scale behavior of disks, we make this microturbulent approximation and adopt $p=\rho \sigma^2$.

Given this decomposition of the energy, we can simplify equation (\ref{totenergy2}) by dropping small terms. For a rotation-dominated disk, $v_r \ll v_\phi$. For a disk with dimensionless viscosity $\alpha$ and scale height $H$, the radial velocity $v_r \sim \alpha (H/r) \sigma$. Thus $v_r \ll \sigma$ unless the disk is thick ($H/r\sim 1$) and accretion happens on a dynamical timescale ($\alpha\sim 1$). For this reason we drop the $v_r^2$ term. We retain terms of order $\sigma$ compared to those of order $v_\phi$, since we are interested in the change in a term of order $\sigma$. Doing so reduces equation (\ref{totenergy2}) to
\begin{equation}
\frac{\partial}{\partial t} \Sigma \left(\frac{v_{\phi}^2}{2} + \frac{3}{2}\sigma^2 + \psi \right)
+ \frac{1}{r}\frac{\partial}{\partial r}\left[r \Sigma v_r \left(\frac{v_{\phi}^2}{2}+\frac{5}{2}\sigma^2 +\psi \right)\right]
=  \Sigma \frac{\partial \psi}{\partial t} +  \frac{1}{2 \pi r}\frac{\partial}{\partial r}(\Omega \calT)
- \dot{\Sigma}_* \left(\frac{v_{\phi}^2}{2} + \frac{3}{2}\sigma^2 + \psi\right) 
+ \calG - \calL,
\label{totenergy}
\end{equation}
where we have rewritten the pressure as $p = \rho \sigma^2$. We can simplify this greatly by using the continuity equation (\ref{continuity2d}) to evaluate the terms on the left-hand side. Doing so, we obtain
\begin{equation}
\frac{1}{2}\Sigma \left[\frac{\partial}{\partial t}\left(v_{\phi}^2+3\sigma^2\right) + v_r\frac{\partial}{\partial r}\left(v_{\phi}^2+3\sigma^2+2\psi\right)\right] + \frac{1}{r} \frac{\partial}{\partial r} \left(r \Sigma v_r \sigma^2\right) = 
\frac{1}{2 \pi r}\frac{\partial}{\partial r}(\Omega \calT) + \calG - \calL,
\end{equation}
which is equation (\ref{eq:econs}).

\section{Solution of the Torque Equation Near the Singularity at the Origin}
\label{innerbc}

Here we obtain the solution to the torque equation (\ref{torquenondim}) near $x=0$ for constant $\beta$ by means of series expansion. Since the nature of the singularity depends on $\beta$, we must handle individual values of $\beta$ separately.

\subsection{Flat Rotation Curve $(\beta = 0)$}

For $\beta=0$ the torque equation (\ref{torquenondim}) reduces to
\begin{equation}
\tau'' - \left(\frac{5}{3 f_g - 1}\right) \frac{s'}{s} \tau' - \frac{1}{(3f_g-1) s^2 x^2} \tau =  \left(\frac{2^{3/2} f_g}{3 f_g-1}\right) \left(\frac{\eta}{\chi}\right) \frac{1}{x}.
\end{equation}
We expand $\tau$ in a power series about $x=0$, $\tau = \sum_{n=0}^{\infty} a_n x^n$, to obtain
\begin{equation}
-\frac{a_0}{(3f_g-1)s^2} x^{-2} - \left[\frac{2\sqrt{2} s^3 f_g \eta+a_1\chi}{(3f_g-1)s^2\chi}\right]x^{-1} - \left\{\frac{a_2[1-2(3f_g-1)s^2]+5 a_1 f_g s s'}{(3f_g-1)s^2}\right\} + O(x) = 0
\end{equation}
Thus the leading coefficients near $x=0$ are
\begin{eqnarray}
a_0 & = & 0 \\
a_1 & = & -2\sqrt{2} s^3 f_g \frac{\eta}{\chi} \\
a_2 & = & \left[\frac{10\sqrt{2} f_g s^4 s'}{1-2(3f_g-1) s^2}\right] \frac{\eta}{\chi}.
\end{eqnarray}

\subsection{Keplerian Rotation Curve $(\beta=-1/2)$}

For $\beta=-1/2$ the torque equation (\ref{torquenondim}) reduces to
\begin{equation}
\label{torquekepler}
\tau'' + \frac{(3f_g-1) s - 10 x s'}{(3f_g-1) s x} \tau' - \frac{3}{4(3f_g-1) s^2 x^3} \tau =  \left(\frac{2^{3/2}f_g s}{3 f_g-1}\right) \left(\frac{\eta}{\chi}\right) \frac{1}{x^{5/2}}.
\end{equation}
We expand $\tau$ in a power series about $x=0$, $\tau = x^{1/2}\sum_{n=0}^{\infty}  a_n x^n$, to obtain
\begin{equation}
-\frac{4 s^3 \eta f_g + 3 a_0 \chi}{4 (3f_g-1) s^2 \chi}x^{-5/2}-\frac{3 a_1}{4(3f_g-1) s^2}x^{-3/2}
- \frac{3 a_2 + 2 s[5 a_0 s' + 3 (3f_g-1) a_1]}{4 (3f_g - 1) s^2} x^{-1/2} + O(x^{1/2}) = 0.
\end{equation}
Thus the leading coefficients near $x=0$ are
\begin{eqnarray}
a_ 0 & = & -\frac{4}{3} s^3 \frac{\eta f_g}{\chi} \\
a_1 & = & 0 \\
a_2 & = & \left(\frac{40 f_g s^4 s'}{9}\right) \frac{\eta}{\chi}.
\end{eqnarray}

\section{Numerical Algorithm for Time-Dependent Disks}
\label{updatealgorithm}

Here we describe our algorithm for numerical solution of the evolution equations (\ref{torquenondim}) and (\ref{sevol}) for time-dependent disks. Let $s_i^{(n)}$ be the velocity dispersion at time $n$ at the center of cell $i$, where $i$ runs from 1 to $N_x$. Cell centers are located at positions $x_i = x_0^{1-(i-1)/(N_x-1)}$, so that the cell spacing in $\ln x$ has a uniform value $d \ln x=-(\ln x_0)/(N_x-1)$. At each time step we obtain the new velocity dispersions $s_i^{(n+1)}$ using an operator-splitting method in which we treat the updates due to the torque explicitly and then perform an implicit diffusion step to suppress spurious numerical oscillations. The explicit part of the algorithm, which occurs first in every time step, is:
\begin{enumerate}
\item We compute the spatial derivatives $(\partial s/\partial x)_i^{(n)}$ at the center of every grid cell using a minmod slope limiter.
\item Using $s_i^{(n)}$ and $(\partial s/\partial x)_i^{(n)}$, we solve the torque equation (\ref{torquenondim}) for $\beta=0$ and the specified values of $\chi$ and $f_g$, using the boundary conditions $\tau'=-x_0$ at $x=x_0$ and $\tau' = -1$ at $x=1$. We solve the equation using the method of shooting to a fitting point, with the fitting point chosen in the middle of the computational grid. We use an adaptive error control method to maintain accuracy, which is necessary because the torque equation can be extremely stiff when $\chi$ is small or $s$ is far from the equilibrium solution. This stiffness also limits the smallest value of $x_0$ we can use and maintain numerical stability at reasonable computational cost.
\item We compute the time derivatives $(\partial s/\partial t)_i^{(n)}$ in each cell using equation (\ref{sevol}). As with $s$, we evaluate the derivatives of $\tau$ using a minmod slope limiter.
\item We set the time step $dt = t^{(n+1)}-t^{(n)}$ to $dt = 0.02 \min_i [|s_i^{(n)} / (\partial s/\partial x)_i^{(n)}|]$.
\item We set $s_i^{(n_*)} = s_i^{(n)}+dt (\partial s/\partial t)_i^{(n)}$, thereby updating $s$ to time $n_*$ for the non-diffusive part of the evolution.
\end{enumerate}

For the next step, we wish to diffuse the velocity dispersion to prevent the development of grid-scale numerical oscillations. In order to guarantee energy conservation, we diffuse the kinetic energy rather than diffusing $s$ directly. We define the dimensionless kinetic energy per unit area in a computational cell by 
\begin{equation}
\label{kedefn}
k = \frac{3}{2}S s^2 = \left(\frac{3 f_g}{\sqrt{2} \pi \chi}\right) \frac{s^3}{x},
\end{equation}
and we evolve this following
\begin{equation}
\label{kdiff}
\frac{\partial k}{\partial t} = \kappa_{\rm diff} \nabla^2 k = \frac{\kappa_{\rm diff}}{x^2} \frac{\partial^2 k}{\partial \ln x^2},
\end{equation}
where $\kappa_{\rm diff}$ is the diffusion coefficient, and in the second step we have evaluated the $\nabla^2$ operator on our cylindrical logarithmic grid. Provided that we set $\partial k/\partial \ln x = 0$ at our inner and outer boundaries, evolution under this equation does not change the total amount of kinetic energy on the grid. We discretize equation (\ref{kdiff}) using centered spatial differences and fully implicit temporal differences:
\begin{equation}
\frac{k_i^{(n+1)} - k_i^{(n_*)}}{dt} = \frac{\kappa_{\rm diff}}{x_i^2} \left[\frac{k_{i+1}^{(n+1)} + k_{i-1}^{(n+1)} - 2 k_i^{(n+1)}}{(d\ln x)^2}\right],
\end{equation}
with the boundary conditions that $k_0^{(n+1)} = k_1^{(n+1)}$ and $k_{N_x+1}^{(n+1)} = k_{N_x}^{(n+1)}$. Rewriting this in matrix form, we have 
\begin{equation}
\label{diffmat}
\mathbf{M}\cdot \mathbf{k}^{(n+1)} = \mathbf{k}^{(n_*)},
\end{equation}
where $\mathbf{k}^{(n+1)}$ and $\mathbf{k}^{(n_*)}$ are the vectors of $k_i$ values at times $n+1$ and $n_*$, respectively, and the matrix $\mathbf{M}$ has elements
\begin{equation}
M_{ij} = \delta_{ij} +
\frac{\kappa_{\rm diff} dt}{(d\ln x)^2}
\left(2\frac{\delta_{ij}}{x_i^2} - \frac{\delta_{i,j-1}}{x_i^2} - \frac{\delta_{i,j+1}}{x_j^2} - \frac{\delta_{i,1}\delta_{j,1}}{x_1^2} - \frac{\delta_{i,N_x}\delta_{j,N_x}}{x_{N_x}^2}\right).
\end{equation}
%\begin{equation}
%M_{ij} =
%\delta_{ij} + \frac{\kappa_{\rm diff} dt}{(d\ln x)^2} \cdot
%\left\{
%\begin{array}{ll}
%x_1^{-2}, & i=j=1\\
%2 x_i^{-2}, & i=j=2\ldots N_x-1 \\
%x_{N_x}^{-2}, & i=j=N_x\\
%-x_i^{-2}, & i=j-1\\
%-x_j^{-2}, & i=j+1\\
%0, & i>j+1 \mbox{ or } i<j-1
%\end{array}
%\right..
%\end{equation}
Since $\mathbf{M}$ is tridiagonal matrix, it is easy to solve equation (\ref{diffmat}) exactly. Thus to take our diffusion step we simply compute $\mathbf{k}^{(n_*)}$ from $\mathbf{s}^{(n_*)}$ (equation \ref{kedefn}), solve equation (\ref{diffmat}) for $\mathbf{k}^{(n+1)}$, and use this to compute $\mathbf{s}^{(n+1)}$, thus completing the timestep. We find that $\kappa_{\rm diff}=0.005$ (for $\chi=0.01$) or $\kappa_{\rm diff}=0.01$ (for $\chi=0.1$) is sufficient to suppress numerical oscillations without significantly changing the solution.

\end{appendix}

\bibliographystyle{apj}
\bibliography{refs}

\end{document}